\documentclass[aps,superscriptaddress]{revtex4}
\usepackage[intlimits]{amsmath}
\usepackage{amssymb}
\usepackage{graphicx}
\usepackage{booktabs}
\usepackage{mathrsfs,slashed}
\usepackage{epsfig}
\usepackage{color}
\usepackage{stix}
\newcommand{\bea}{\begin{eqnarray}}
\newcommand{\eea}{\end{eqnarray}}

\begin{document}
\title{Generalized Beth--Uhlenbeck entropy formula from the $\Phi-$derivable approach}

\author{D.~Blaschke}
\email{blaschke@ift.uni.wroc.pl}
\affiliation{Institute of Theoretical Physics, 
	University of Wroclaw, 
	Max Born place 9, 
	50-204 Wroclaw, Poland}
\affiliation{Helmholtz-Zentrum Dresden-Rossendorf (HZDR), Bautzner Landstrasse 400, 01328 Dresden, Germany}
\affiliation{Center for Advanced Systems Understanding (CASUS), Untermarkt 20, 02826 G\"orlitz, Germany}
%\affiliation{Laboratory for Theoretical Physics, Joint Institute for Nuclear Research, Joliot-Curie street 6, 141980 Dubna, Russia}
%\affiliation{National Research Nuclear University (MEPhI), Kashirskoe Shosse 31, 115409 Moscow, Russia}
\author{G.~R\"opke}
\email{roepke@uni-rostock.de}
\affiliation{Institut f\"ur Physik, 
	Universit\"at Rostock, 
	Albert-Einstein-Strasse 23-24, 
	18059 Rostock, Germany}
%\affiliation{National Research Nuclear University (MEPhI), Kashirskoe Shosse 31, 115409 Moscow, Russia}
\author{G.~Baym}
\email{gbaym@illinois.edu }
\affiliation{Department of Physics, 
	University of Illinois, 
	1110 W. Green Street, 
	Urbana, IL 61801, USA}

%%%%%%%%%%%%%%%%%%%%%%%%%%%%%%%%%%%
\date{\today}
\begin{abstract}
We derive a generalized Beth-Uhlenbeck formula for the entropy 
%as well as the density, 
of a dense fermion system with strong two-particle correlations,  including scattering states and bound states.  We work within  the $\Phi-$derivable approach to the thermodynamic potential.   The formula takes the form of an energy-momentum integral over a statistical distribution function times a unique spectral density.   In the near mass-shell limit,  the spectral density reduces, contrary to na\"{i}ve expectations,  not to a Lorentzian but rather to a  "squared Lorentzian" shape.   
The relation of the Beth-Uhlenbeck formula to the $\Phi$-derivable approach is exact at the two-loop level for $\Phi$.    
The formalism we  develop, which extends the Beth-Uhlenbeck approach beyond the low-density limit,  includes Mott dissociation of bound states, in accordance with Levinson's theorem, and the self-consistent back reaction of correlations in the fermion propagation.  We discuss applications to further systems, such as quark matter and nuclear matter.
\end{abstract}
%\begin{keyword}
%Generalized Beth-Uhlenbeck formula, $\Phi-$ derivable approach, Levinson theorem, Mott effect
%\end{keyword}
\maketitle
%\tableofcontents
\section{Introduction}\label{sec1}

Fermionic systems that can form bound states are very common in physics.
Well known examples include plasmas, in particular the hydrogen plasma and the electron-hole plasma in semiconductors, ruled by the Coulomb interaction.
Other examples are nuclear matter and the quark-gluon plasma, 
where the constituents interact via the strong QCD interaction.
All these systems share in common the formation of bound states and their dissolution at high densities.
Although treating the correlations in such interacting quantum systems is very challenging, they can be treated by similar approaches within the  framework of quantum statistical mechanics.

Various methods have been worked out in quantum statistics to treat 
many-particle systems, such as ionic plasmas and warm dense matter, the 
electron-hole plasma in semiconductors, the quark-gluon plasma, nuclear 
matter, etc. In particular, the Matsubara Green's function method and the 
use of Feynman diagrams provide the tools to investigate 
these interacting many-particle systems. Partial summations can be
performed to introduce quasiparticles and self-energies, screening of 
interactions, vertex corrections, formation of bound states, and 
other collective excitations, leading to an improved perturbative approach and successful approximation schemes. 
See  the book of 
%Kadanoff and Baym 
\cite{KB} for the quantum statistical approach to the Coulomb many-particle problem via Green's functions methods in equilibrium and nonequilibrium.

As pointed out by %Baym and Kadanoff  
\cite{Baym:1961zz,Baym:1962sx} and others, the selection of diagrams that define special approximations must be taken with care. 
%\redflag
{In particular, one has to obey the conservation laws and avoid double counting of contributions. To obey the Ward identities, which assure that for given self-energy contributions, appropriate vertex corrections are included in order to fulfill exact sum rules. } 
As quasiparticles become shifted and broadened, one has to take into account effects that contribute in the same order of 
magnitude because of possible compensation effects.
Although of high interest, it is a big challenge in 
linear response theory and kinetic theory to describe the evolution of many-particle systems while guaranteeing the conservation of particle number, energy, and momentum.  
The $\Phi$-functional approach provides a prescription \citep{Baym:1962sx}
to construct such approximations.   
This concept of conserving approximations has been further elaborated by, e.g., %Riedel 
\cite{Riedel:1968}, %Vanderheyden and Baym 
\cite{Vanderheyden:1998ph}, %Blaizot et al. 
\cite{Blaizot:1999ap}, \cite{Blaizot:1999ip} and \cite{Blaizot:2000fc}.	

Another earlier approach to the thermodynamic properties of non-relativistic plasmas was worked out 
using the Beth-Uhlenbeck equation for the second virial coefficient \citep{Beth:1936zz,Beth:1937zz}.
To treat higher densities, this low-density expansion requires improvements,
which can be done by introducing quasiparticles.
%Zimmermann and Stolz 
\cite{Zimmermann:1985ji} showed then that the contribution of the scattering states 
must be modified, and a $\sin^2\delta-$ term appears in the Beth-Uhlenbeck formula to compensate the mean-field contributions, which are already taken into account in the quasiparticle shift.
A detailed discussion of this generalized Beth-Uhlenbeck formula is found in %Schmidt et al. 
\cite{Schmidt:1990oyr}.

Our aim in this paper is to show the equivalence of the $\Phi$-functional 
%\redflag
{at the two-loop level with}
the generalized Beth-Uhlenbeck approaches.  We use a simple model of interaction for which we construct the spectral functions, and
with the help of generalized optical theorems we show that the appearance of the $\sin^2\delta-$ term
corresponds to the squared Breit-Wigner profile in the spectral function.

We follow the approach of %Vanderheyden and Baym 
\cite{Vanderheyden:1998ph} who start from relativistic field theories described by the QED Lagrangian. Other systems where these methods are of interest include
Fermi liquids such as $^3$He, as considered by %Riedel 
\cite{Riedel:1968}, and the quark-gluon plasma, see %Blaizot et al. 
\cite{Blaizot:2000fc}.
Our approach, demonstrated in a simple case, can be generalized to more complex systems.

We mention here related examples that show the necessity of using consistent approximations in plasma physics. %References~
\cite{RKKKZ78} and \cite{ZKKKR78}, following the Green's function approach, derived an effective in-medium Schr\"odinger equation for plasmas and showed that when the Fock shift of single-particle energies are considered, one also has to take Pauli blocking of the interaction into account.  When introducing screening for the interaction potential, one has also to 
include the Montroll-Ward self-energy terms (the rainbow diagrams); and when improving the two-particle propagators in the polarisation loop by medium effects, one must also improve the vertex \citep{RD79}. 

In addition to plasma physics, associated work has been carried out for 
%\redflag
{strongly interacting matter. 
Applications of the Luttinger-Ward formalism \citep{Luttinger:1960ua} of the $\Phi$-functional to nuclear matter have been recently summarized in \cite{Rios:2007}, with special applications to the entropy of correlated nuclear matter 
\citep{Rios:2006nr}. 
However, these approaches do not make use of phase shifts to describe the continuum states of the two- or many-nucleon scattering processes so that their relation with the Beth-Uhlenbeck description is not obvious. 
In the cases where free space scattering phase shifts are experimentally measured, the standard Beth-Uhlenbeck approach has been used;  following the seminal work of Dashen, Ma and Bernstein \citep{Dashen:1969ep} this method is called the S-matrix approach. A few examples, including  very recent works, are:
\cite{Welke:1990za,Venugopalan:1992hy,Weinhold:1997ig,Horowitz:2005nd,Mallik:2008zt,Lo:2017ldt,Lo:2017sde,Vovchenko:2017drx,
Andronic:2018qqt,Lo:2019who,Donigus:2022haq} and \cite{Honek:2026exd}. 
These approaches use hadron-hadron scattering phase shifts without medium modifications and thus their application becomes questionable in situations at high densities and temperatures, where chiral symmetry restoration sets in and hadrons dissociate into their quark constituents.    

A Beth-Uhlenbeck approach with hadrons as bound states of the quark-quark scattering has been developed within the NJL model description of chiral quark-meson matter \cite{Hufner:1994ma,Zhuang:1994dw,Wergieluk:2012gd,Yamazaki:2012ux,Blaschke:2013zaa,Blaschke:2014zsa}, where the Mott dissociation of hadrons is included in the model phase shifts as a dramatic medium effect. 
In these approaches, however, the thermodynamic functions are given in the standard Beth-Uhlenbeck form, without the $\sin^2\delta-$ term.

There are only a few examples where the generalized Beth-Uhlenbeck approach with the $\sin^2\delta-$ term has been used for strongly interacting matter
\cite{Schmidt:1990oyr,Blaschke:2016fdh,Ropke:2020fut,Blaschke:2023pqd}.}
%A formulation of the back-reaction problem within the path integral approach has been given for the NJL model \citep{Blaschke:2017boi} where it was shown how the form of a $\Phi-$derivable thermodynamical potential could be obtained for quark matter with composite mesons and how the diagrammatic expansion for the  $\Phi-$functional is related to the $1/N_c$ expansion of the NJL model.
%\redflag
{
Only the generalized form of the Beth-Uhlenbeck approach with the $\sin^2\delta-$ term makes it equivalent to the $\Phi-$derivable approach in the two-loop approximation as we will show in the present work.
}

%%%%%%%%%%%%%%%%%%%%%%%%%%%%%%%%%%%%%%%%%%%%%%%%%%%%
\section{$\Phi-$derivable approach to the electron-positron plasma in QED in the two-loop approximation}
%%%%%%%%%%%%%%%%%%%%%%%%%%%%%%%%%%%%%%%%%%%%%%%%%%%%
A fundamental approach to the theory of plasmas starts from quantum electrodynamics (QED), with fermions coupled to the electromagnetic field, the approach of %Vanderheyden and Baym 
\cite{Vanderheyden:1998ph} in discussing self-consistent approximations in relativistic plasmas.   Methods similar to those proposed in that paper can be applied to other plasmas, e.g. the electron-proton plasma. 
%\subsection{Fermion-boson theories of Yukawa type}

\subsection{Thermodynamical potential for the electron-positron-photon system 
with a ``sunset" $\Phi-$functional}

Applying the $\Phi-$derivable formulation \cite{Baym:1962sx} to QED thermodynamics, we closely follow %Ref. 
\cite{Vanderheyden:1998ph} where
the thermodynamic potential reads
\begin{eqnarray}
	\beta \Omega(T,\mu)
	= \Phi\left[G, D \right] - {\rm Tr} \Sigma G 
    + {\rm Tr} \ln\left(-\gamma_0 G\right)
    + \frac{1}{2} {\rm Tr} \Pi D 
    -  \frac{1}{2} {\rm Tr}\ln\left(-D\right),
\label{eq:OmegaPhi}
\end{eqnarray}	
where $G$ is the electron Green's function with self-energy $\Sigma$, and $D$ is the photon Green's function with self-energy or polarization operator $\Pi$; $\gamma_0$ is the Dirac matrix,
%\redflag
{$\beta=1/T$ the inverse temperature and $\mu$ the electron chemical potential}.   
The functional $\Phi\left[G, D \right]$ is the sum of all skeleton diagrams of the thermodynamic potential. This ansatz guarantees that the stationary condition for the thermodyanamic potential against variations of the Green's functions,
\begin{equation}
    \frac{\delta \Omega}{\delta G}=\frac{\delta \Omega}{\delta D}=0~,
\label{eq:stationarity}
\end{equation}
is equivalent to the statement that the 
propagators $G$ for electrons and $D$ for photons fulfill the Dyson equations
\begin{eqnarray}
\label{eq:Dyson}
    G^{-1}&=&G_0^{-1}-\Sigma~, \nonumber\\
    D^{-1}&=&D_0^{-1}-\Pi~,
\end{eqnarray}
where $G_0$ and $D_0$ are the bare electron and photon propagators, and  $\Sigma$ and $\Pi$ are calculated as functional derivatives,
\begin{equation}
    \Sigma=\frac{\delta \Phi}{\delta G}
    ~,~~\Pi=2\frac{\delta \Phi}{\delta D}~.
\label{eq:Self}
\end{equation}
We assume the Coulomb gauge.

Using the spectral representations for the Green's functions, Eq. (\ref{eq:Dyson}),
\begin{eqnarray}
\label{spectral}
    G(\omega_n,{\bf p}) &=& \int_{-\infty}^{\infty} \frac{d\omega}{2\pi} \frac{A(\omega,{\bf p})}{\omega_n - \omega}~,\\
    D_T(\omega_n,{\bf q}) &=& \int_{-\infty}^{\infty} \frac{d\omega}{2\pi} \frac{B_T(\omega,{\bf q})}{\omega_n - \omega}~,
\label{spectral-T}    
\\
    D_L(\omega_n,{\bf q}) &=& \int_{-\infty}^{\infty} \frac{d\omega}{2\pi} \frac{B_L(\omega,{\bf q})}{\omega_n - \omega}~+~\frac{1}{{\bf q}^2}~,
\label{spectral-L}    
\end{eqnarray}
with the fermionic Matsubara frequencies 
$\omega_n=(2n+1)\pi i\,T + \mu$ for electrons and bosonic Matsubara frequencies $\omega_n= 2\pi i\, n \,T$ for photons. 

  The frequency sums in (\ref{eq:OmegaPhi})
can be evaluated by contour integration and yield
\begin{eqnarray}
	\Omega(T,\mu)
	&=& T \Phi\left[G, D \right] 
    + \sum_{\bf p}\int_{-\infty}^{\infty} \frac{d\omega}{2\pi} f(\omega) 
    {\rm Tr}_{\rm Dirac} {\rm Im} \left[ \Sigma G 
    + \ln\left(-\gamma_0 G^{-1}\right)\right]
    \nonumber\\
    &&+ \sum_{\bf q} 
%    \intbar 
    \int \hspace{-3mm}-\hspace{2mm}
    \frac{d\omega}{2\pi}\,
     n(\omega) \sum_{{l}=L,T} g_l {\rm Im} \left[ \Pi_l D_l 
    + \ln\left(-D_l^{-1}\right)\right],
\label{eq:Omega}
\end{eqnarray}	
with the statistical factors 
$f(\omega)=\{\exp[\beta(\omega-\mu)]+1\}^{-1}$ 
for fermions and
$n(\omega)=\{\exp(\beta\omega)-1\}^{-1}$ 
for bosons.  (The slash through the boson integral indicates the principal value.)   The degeneracy factors are $g_L=1$  for longitudinal modes and $g_T=2$ for transverse modes.

\begin{figure}[htb]
%\centerline{
\parbox{0.05\textwidth} {
$\Phi=\frac{1}{2}$
}%\hfill
\parbox{0.08\textwidth}{
\includegraphics[width=0.08\textwidth]{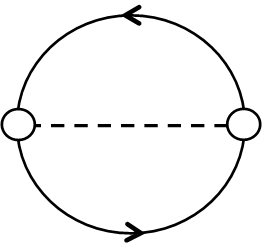}
}
%\parbox{0.02\textwidth}{$+\frac{1}{2}$}
%\parbox{0.08\textwidth}{
%\includegraphics[width=0.08\textwidth]{Fig5A}
%}
%\parbox{0.01\textwidth}{$+$%\frac{1}{2}$
%}
%\parbox{0.08\textwidth}{
%\includegraphics[width=0.08\textwidth]{Fig5B}
%}
%\parbox{0.02\textwidth}{$+\frac{1}{2}$}
%\parbox{0.08\textwidth}{
%\includegraphics[width=0.08\textwidth]{Fig5G}
%}
\caption{
The two-loop functional $\Phi$, where the solid line is the electron propagator and the dashed line the photon propagator.}  
\label{fig:Phi}
\end{figure}

From this expression for the thermodynamic potential all other equations of state can be obtained by differentiation.   
The entropy $S$ and particle number $N$ are obtained as derivatives with respect to the temperature and the chemical potential:
\begin{equation}
    S=-\frac{\partial \Omega(T,\mu)}{\partial T}\bigg|_\mu,~~
    N=-\frac{\partial \Omega(T,\mu)}{\partial \mu}\bigg|_T .
    \label{sn}
\end{equation}
Because of the stationarity \eqref{eq:stationarity}, we can ignore the spectral densities of the propagators when differentiating  
$\Phi\left[G, D \right]$, and need only differentiate the statistical factors. 
Let us consider the entropy as the derivative with respect to the temperature,
\begin{equation}
    S=-\frac{\partial \Omega(T,\mu)}{\partial T}\bigg|_{\mu,V,A,B}
    = S_f+S_b+S^{\prime},
\end{equation}
with the fermionic and bosonic contributions,
\begin{eqnarray}
   \label{eq:Sf}
    S_f&\equiv&-\sum_{\bf p}\int_{-\infty}^{\infty} \frac{d\omega}{2\pi} \frac{\partial f(\omega)}{\partial T} 
    {\rm Tr}_{\rm Dirac}  \left[{\rm Im} \,\Sigma \,\,{\rm Re}\, G 
    + {\rm Im} \ln\left(-\gamma_0 G^{-1}\right)\right], \\
    S_b&\equiv& -\sum_{{\bf q},l} g_l 
%    \intbar 
    \int \hspace{-3mm}-\hspace{2mm}
    \frac{d\omega}{2\pi}\, \frac{\partial n(\omega)}{\partial T}  
     \left[ {\rm Im}\, \Pi_l \,\,{\rm Re}\, D_l 
    + {\rm Im} \ln\left(-D_l^{-1}\right)\right],
    \label{eq:Sb}
\end{eqnarray}
respectively. 
The remaining term, % $S^\prime$ vanishes
\begin{eqnarray}
\label{eq:Sprime}
    S^\prime &\equiv& 
    -\frac{\partial(T \Phi)}{\partial T}\bigg|_{A,B}
    - \sum_{\bf p}\int_{-\infty}^{\infty} \frac{d\omega}{2\pi} \frac{\partial f(\omega)}{\partial T} 
    {\rm Tr}_{\rm Dirac}  \left[{\rm Im} \,G \,\,{\rm Re}\, \Sigma \right]
%    \nonumber\\   &&
    -\sum_{{\bf q},l} g_l 
    %\intbar 
    \int \hspace{-3mm}-\hspace{2mm}
    \frac{d\omega}{2\pi}\, \frac{\partial n(\omega)}{\partial T} {\rm Im}\, D_l\,\, {\rm Re} \,\Pi_l 
    %= 0 
    ~,
\end{eqnarray}
has the important property that it vanishes in the two-loop approximation for $\Phi$, as illustrated diagrammatically in Fig. \ref{fig:Phi} for fermions (solid lines) interacting via the exchange of a boson (dashed line). 
(In~\cite{Vanderheyden:1998ph} the same diagram is denoted as ``one-loop.") 
This vanishing has been shown for QED in~\cite{Vanderheyden:1998ph} and for QCD in \cite{Blaizot:2000fc}. 
It is based on the cancellation of the derivative of the $\Phi-$functional and the terms with the trace over the product of the imaginary part of the propagator (spectral density) of a particle and the real part of its self-energy, see Eq. \eqref{eq:Sprime}. The cancellation holds also for both the entropy and particle density in the 
two-loop approximation, $S^\prime = N^\prime = 0$.   
We demonstrate this cancellation for the simple case of a scalar field $\phi$ with a $\phi^3$ interaction vertex in Appendix \ref{app:warmup}.

\begin{figure}[htb]
%\centerline{
\parbox{0.05\textwidth} {
$\Sigma=$
}%\hfill
\parbox{0.08\textwidth}{
\includegraphics[width=0.08\textwidth]{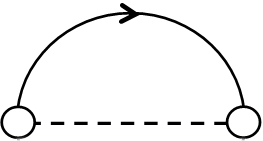}
}%\hfill
\hspace{1cm}
\parbox{0.05\textwidth}{$\Pi=$}
\parbox{0.08\textwidth}{
\includegraphics[width=0.08\textwidth]{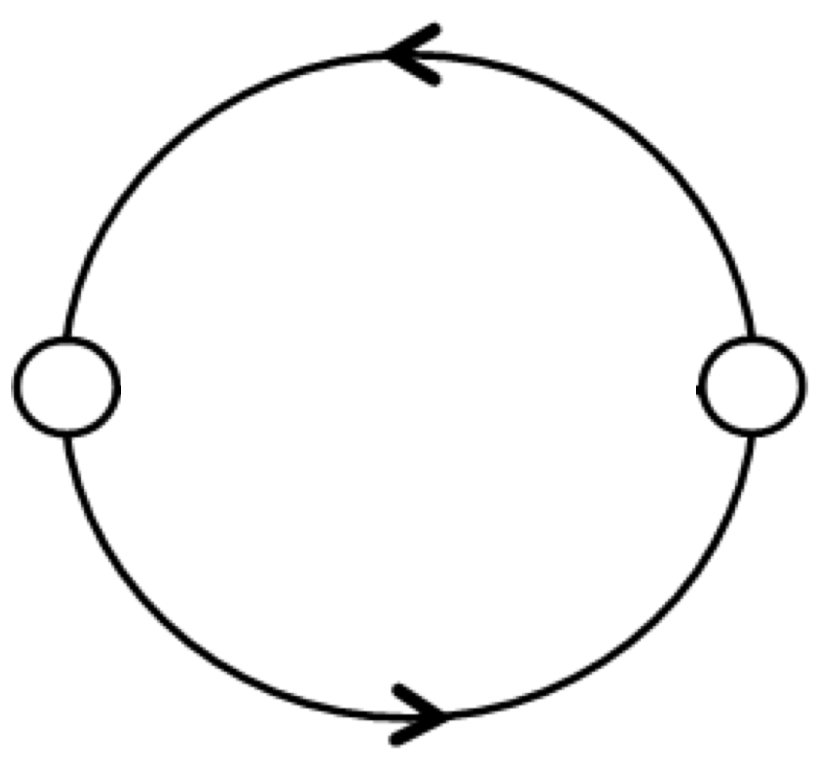}
}
%\parbox{0.01\textwidth}{$+$%\frac{1}{2}$
%}
%\parbox{0.08\textwidth}{
%\includegraphics[width=0.08\textwidth]{Fig5B}
%}
%\parbox{0.02\textwidth}{$+\frac{1}{2}$}
%\parbox{0.08\textwidth}{
%\includegraphics[width=0.08\textwidth]{Fig5G}
%}
\caption{
Self-energy diagrams derived from the $\Phi$ in Fig. \ref{fig:Phi}.} 
\label{fig:Self}
\end{figure}

Figure~\ref{fig:Self} shows the corresponding self-energy diagrams obtained as variational derivatives,  Eqs. \eqref{eq:Self}. 
%

% \redflag{NOTE:  THE FILE SENT ON 22 NOV 25 HAD EDITS ONLY UP TO HERE!!!}
 
In the following we focus on the bosonic entropy, $S_b$, and rewrite Eq.~\eqref{eq:Sb} using the identity
\begin{equation}
    \frac{\partial n(\omega)}{\partial T} = - \frac{\partial \sigma_n(\omega)}{\partial \omega} ~,
\end{equation}
where 
\begin{equation}
\label{sigman}
\sigma_n(\omega)\equiv -n(\omega) \ln n(\omega) + (1+n(\omega))\ln(1+n(\omega))
\end{equation}
is the contribution to the  entropy of  a photon mode of frequency $\omega$.   
After partial integration over $\omega$ one obtains
\begin{equation}
    S_b = \sum_{{\bf q}} 
%    \intbar_0^\infty 
    \int_0^\infty \hspace{-5.5mm}-\hspace{3mm}
    \frac{d\omega}{2\pi}\, \sigma_n(\omega)B_s(\omega,{\bf q}),
\label{eq:Sb2}
\end{equation}
where 
\begin{eqnarray}
    B_s(\omega,{\bf q})&=&\sum_{l=L,T} g_l B_{s,l}(\omega,{\bf q}),
\end{eqnarray}
 with
\begin{eqnarray}
    B_{s,l}(\omega,{\bf q})&=&
    \frac{\partial}{\partial \omega}\left[{\rm Re} D_lL_l - 2 {\rm Im} \ln (-D_l^{-1}) \right],
    \label{eq:Bs}
\end{eqnarray}
and $L_l(\omega,{\bf q})\equiv -2\, {\rm Im}\, \Pi_l(\omega+i 0^+,{\bf q})$ the imaginary part of the photon self-energy.

 {
To interpret the contents of this formula, we expand the inverse propagator of the Dyson equation \eqref{eq:Dyson} in the vicinity of the plasmon poles $\omega_l$ which are defined by its zeroes, ${\rm Re}\,D_l^{-1}(\omega=\omega_l,{\bf q})=0$, to obtain
${\rm Re}\,D_l^{-1}\simeq(\omega-\omega_l)/Z_l$ and ${\rm Im}\,D_l^{-1}=L_l/2$,
where $Z_l = d {\rm Re} D/d\omega|_{\omega_l}$.
In this approximation the propagator is
\begin{equation}
    D_l=Z_l \frac{(\omega-\omega_l)+i\, Z_l L_l/2}{(\omega-\omega_l)^2+(Z_l L_l/2)^2}.
\end{equation}
The frequency derivatives in Eq. \eqref{eq:Bs} can now be evaluated to give
\begin{eqnarray*}
    \frac{\partial}{\partial \omega}{\rm Re}\, D_l\,L_l&=&\frac{Z_lL_l}{(\omega-\omega_l)^2+(Z_l L_l/2)^2}-\frac{2(\omega-\omega_l)^2L_l}{[(\omega-\omega_l)^2+(Z_l L_l/2)^2]^2},\\
    \frac{\partial}{\partial \omega}{\rm Im} \ln (-D_l^{-1})&=&\frac{Z_lL_l/2}{(\omega-\omega_l)^2+(Z_l L_l/2)^2}.
\end{eqnarray*}
}
As discussed in \cite{Vanderheyden:1998ph},  the entropy spectral function \eqref{eq:Bs} in the vicinity of the longitudinal and transverse plasmon modes of frequencies $\omega_L(q)$ and $\omega_T(q)$, takes the form of a simple squared Lorentzian
\begin{equation}
\label{eq:Lorentz-squared}
    B_{s,l}(\omega_q,{\bf q})\simeq 
    \frac{4(Z_lL_l/2)^3}{[(\omega_q-\omega_l(q))^2+(Z_lL_l/2)^2]^2},
\end{equation}
%  \redflag{VERIFY THIS DEFINITION AND USE A COMMON NOTATION FOR REAL AND IMAGINARY PARTS THROUGHOUT THE PAPER}                   
When compared with the spectral densities of photons $B_l$,
which have a Lorentzian form close to the poles, 
$B_l\simeq (Z_lL_l)/[(\omega_q-\omega_l(q))^2+(Z_lL_l/2)^2]$, we see that the $B_{s,l}$ have a stronger peak and smaller wings. 

 {
The entropy formula \eqref{eq:Sb2} with the spectral function \eqref{eq:Bs} holds for the rather wide class of two-loop diagrams for the $\Phi-$functional (as, e.g., given in Fig. \ref{fig:Phi}) since for these diagrams the cancellation $S^\prime=0$ of Eq. \eqref{eq:Sprime} applies, as it  has been stated in Refs. \cite{Vanderheyden:1998ph,Blaizot:2000fc}.
In the next section we show that the equation of state for the entropy \eqref{eq:Sb2} in this approximation takes the form of a generalized Beth-Uhlenbeck equation.
}

\subsection{Generalized Beth-Uhlenbeck formula}
\label{sec:GBU}

%In the following we denote the real and imaginary parts of complex functions by subscripts $R$ and $I$, respectively, as introduced in Appendix~\ref{app:GOT}. 
 {
In this section, we suppress the mode index $l=L,T$ for brevity and denote the frequency derivative by a prime.
We rewrite the second term on the right side of Eq.~\eqref{eq:Bs} as
\begin{eqnarray}
    \label{corr}
	\left[{\rm Im}\ln(- D^{-1})\right]^\prime
	&=& -{\rm Im}\left(D\Pi^\prime\right)\nonumber\\
	&=&
	\underbrace{({\rm Re}D^\prime\,{\rm Im}\Pi-{\rm Im}D\,{\rm Re}\Pi^\prime)}_{
    2{\rm Im}(D\,{\rm Im}\Pi\, {D^*}^\prime\,{\rm Im}\Pi)} 
	-\underbrace{({\rm Im}\Pi \,{\rm Re}D^\prime+{\rm Re}D\,{\rm Im}\Pi^\prime)}_{({\rm Im}\Pi \,{\rm Re}D)^\prime}~,\nonumber\\
\end{eqnarray}
where we prove the relation 
\begin{equation}
    {\rm Re}D^\prime\,{\rm Im}\Pi-{\rm Im}D\,{\rm Re}\Pi^\prime =
    2{\rm Im}(D\,{\rm Im}\Pi\, {D^*}^\prime\,{\rm Im}\Pi),
    \label{eq:deropt}
\end{equation}
which can be called "derivative optical theorem", in Appendix~\ref{app:GOT}.
The right side of Eq. \eqref{eq:deropt} is similar to the final result for the thermodynamic potential derived by %Dashen, Ma and Bernstein 
\cite{Dashen:1969ep} in their Eq. (4.32). It can be viewed as its generalisation from the low-density limit to the self-consistent quasiparticle level which includes the effects of phase space occupation via $\Pi_I$. For its evaluation in the case of deuterons in nuclear matter, see Ref. \cite{Schmidt:1990oyr}. 
}

We now introduce the phase function $\delta$ of the polar representation of the propagator $D=|D|\exp{(i\delta)}$ 
%and use the optical theorems derived in Appendix~\ref{app:GOT}, 
to find the relationships
\begin{eqnarray}
    D\,{\rm Im}\Pi&=&\sin\delta \,{\rm e}^{i\delta}~~,~~{D^*}^\prime\, {\rm Im}\Pi
= -i\delta^\prime\sin\delta\,{\rm e}^{-i\delta}~, 
\end{eqnarray}
which together with Eq. \eqref{eq:deropt} allow to rewrite Eq.~\eqref{eq:Bs} in 
terms of the phase shift $\delta$ as
\begin{eqnarray}
    \label{sin2}
-\left[{\rm Re}D\, {\rm Im}\Pi + {\rm Im}\ln(- D^{-1})\right]^\prime
%= - 2{\rm Im}(D\Pi_I {D^*}^\prime\Pi_I) 
= 2 \delta^\prime \sin^2\delta
%~~,~~
%\Pi_I S_R=\sin\delta\cos\delta
~.
\end{eqnarray}

Inserting (\ref{sin2}) into \eqref{eq:Bs} we arrive at the main result of this paper,
the generalized Beth-Uhlenbeck equation of state for the entropy \eqref{eq:Sb2}
\begin{eqnarray}
    \label{Sb-GBU}
	S_b &=&
	2V\int\frac{{\rm d}^3q}{(2\pi)^3}~
%	\intbar_{0}^{\infty}
    \int_0^\infty \hspace{-5.5mm}-\hspace{3mm}
    \frac{{\rm d}\omega}{\pi}~
	\sigma_n(\omega)\, \sin^2\delta(\omega,{\bf q})
	\frac{\partial \delta(\omega,{\bf q})}{\partial \omega} 
\end{eqnarray}	
%\redflag
{with $\sigma_n(\omega)$ given by \eqref{sigman}}.
This result  has interesting consequences for the phenomenology of fermion-boson systems.
Similar to the standard Beth-Uhlenbeck equation of state, in the two-loop approximation for  
$\Phi$ one obtains a good quasiparticle picture for the statistical system as a superposition of elementary (e.g., quarks) as well as composite (e.g., hadronic) degrees of freedom which in general are all off-shell, including the elementary ones, as a result of the self-consistency of the approach. 
Ideal quasiparticles with on-mass-shell behaviour are obtained when the corresponding phase function
degenerates to a step function at the mass of the particle.

 {
The spectral function \eqref{sin2} of the generalized Beth-Uhlenbeck formula \eqref{Sb-GBU} has the shape of the "squared Lorentzian"
\eqref{eq:Lorentz-squared} when for the phase shift the resonance shape with the plasmon pole at $\omega=\omega_R$ and the width $\Gamma=Z_R\, {\rm Im}\Pi$ is assumed
\begin{equation}
    \delta(\omega) = - \arctan \left[ \frac{\Gamma}{\omega - \omega_R}\right]~. 
\end{equation}
This form corresponds to a Breit-Wigner (or Lorentz-) profile for the derivative of the phase shift
\begin{equation}
    \label{BW}
\frac{\partial \delta(\omega)}{\partial \omega} = -\frac{\Gamma}{(\omega-\omega_R)^2 + \Gamma^2}~,
\end{equation}
}
then the spectral weight in the generalized Beth-Uhlenbeck formula for the entropy does {\it not} have the Breit-Wigner shape expected from the standard Beth-Uhlenbeck approach, but rather it is a "squared Lorentzian" profile
\begin{equation}
    \sin^2\delta(\omega) \frac{\partial \delta(\omega)}{\partial \omega}
= \frac{\Gamma^3}{[(\omega - \omega_R)^2+ \Gamma^2]^2}~.
\end{equation}
This result, which is obtained here for the simplifying ansatz (\ref{BW}) for the phase shift, has been found
before within the $\Phi-$derivable approach in Ref.~\cite{Vanderheyden:1998ph} for QED (see Eq. \eqref{eq:Lorentz-squared} of the previous subsection) and in 
Ref.~\cite{Weinhold:1998} for the entropy of the Lee model.   
%\redflag{MOVED FROM THE INTRODUCTION:  MERGE IN BETTER. }  
Similar results were obtained also in other approaches and in a different context, for instance 
%\cite{Morozov:2009}. It has also been obtained in a different context 
within the kinetic theory approach to radiation in non-equilibrium plasmas  \cite{Morozov:2009}.
For applications in thermodynamic systems, the effect of considering the second virial coefficient of quasiparticles in the medium rather than free on-shell particles is a sharpening of the spectral distribution describing the resonant scattering (correlation). 

%\redflag
{
\subsection{Two-particle correlations and bound states in a Yukawa-model plasma}
%}
%\redflag{
The simple diagram, Fig. \ref{fig:Phi} is not suitable to describe two-particle correlations. For this purpose we replace the dashed line of 
Fig. \ref{fig:Phi} symbolizing a  gauge boson exchange with a double line that stands for the propagator of a two-particle correlation, here an electron-positron state that we generically call positronium, see Fig. \ref{fig:sunset2}.
Since we have changed the channel and the corresponding Matsubara frequencies, the frequency $\Omega$ is related to the total energy of the two-particle system, and the total momentum $\bf P$ is a conserved quantity.
\begin{figure}[h]
\centerline{\includegraphics[width=0.1\textwidth]{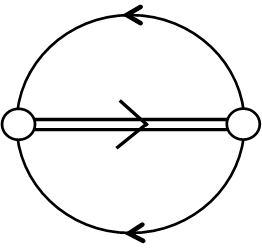}}
\caption{%\redflag
{
Feynman diagram for the $\Phi$-functional for a fermion plasma with two-particle correlations described by the two-particle propagator $G_2(\Omega)$ (double line) at two-loop order (sunset diagram).}
\label{fig:sunset2}}
\end{figure}

The corresponding fermion self-energy is obtained from the 
sunset-type $\Phi$-functional of Fig.~\ref{fig:sunset2} by functional derivation with respect to the fermion propagator and takes the form shown in Fig. \ref{fig:SE-T}. 
\begin{figure}[h]
\centerline{\includegraphics[width=0.25\textwidth]{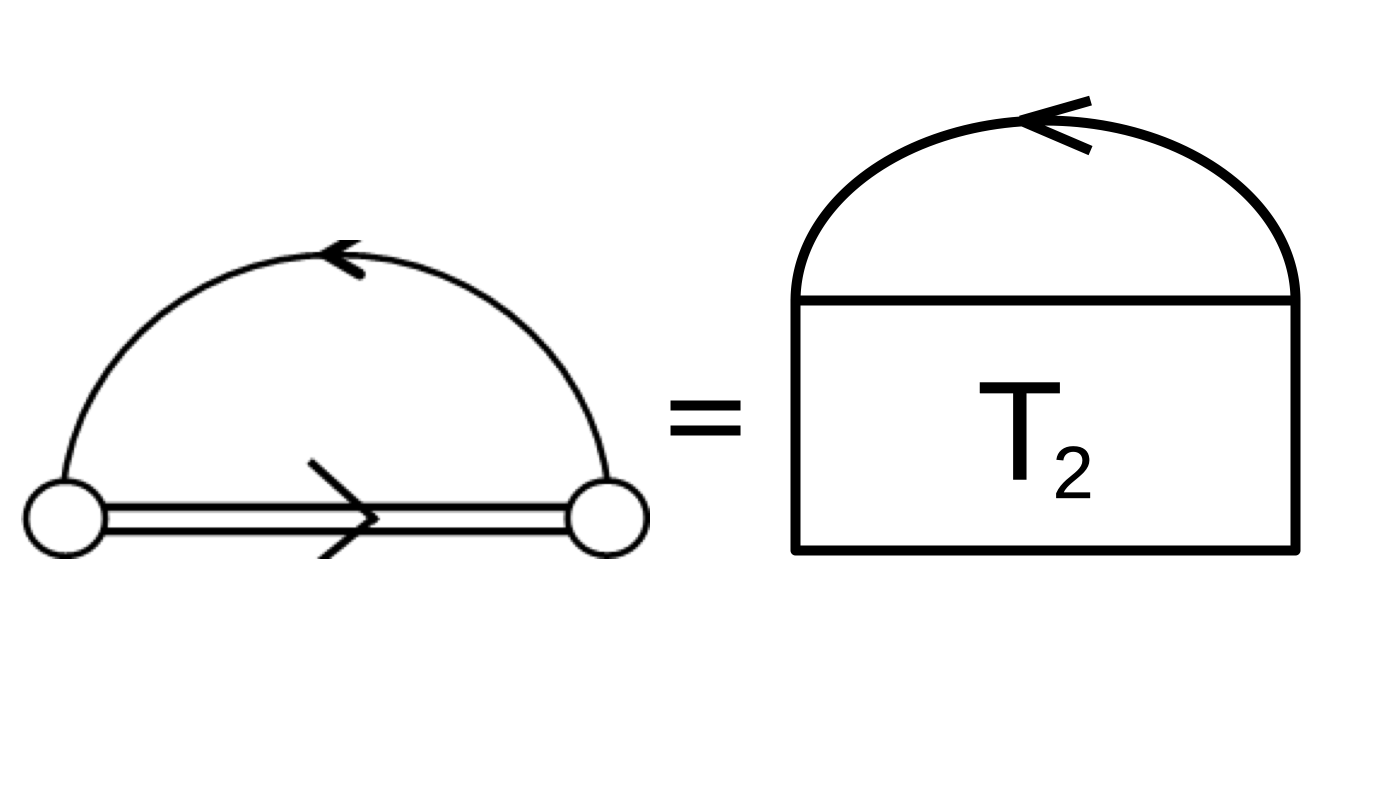}}
\caption{%\redflag
{
Feynman diagram for the fermion self-energy corresponding to the $\Phi-$ functional of Fig. \ref{fig:sunset2} expressed with the two-particle propagator $G_2(\Omega)$ (double line) or, equivalently, with the two-particle T-matrix $T_2(\Omega)$.}
\label{fig:SE-T}}
\end{figure}
The double line stands for the full two-particle propagator which by its analytic properties is equivalent to the two-particle T-matrix $T_2(\Omega)$, see Fig. \ref{fig:SE-T}. 
For applications to strong correlations and bound states, we restrict ourselves to the consideration of an instantaneous interaction potential in the Coulomb gauge and the ladder approximation to the T-matrix, which fulfills the Bethe-Salpeter equation  
\begin{equation}
\label{eq:T}
    T_{2}(\Omega)=V_{\rm D} + V_{\rm D} G_2^0(\Omega) T_{2}(\Omega).
\end{equation}
Here, $G_2^0(\Omega)$ denotes the Matsubara-summed product of two fully dressed one-particle Green's functions, 
\begin{equation}
    G_2^0(\Omega)=\sum_n G(\Omega-\omega_n)G(\omega_n),
\end{equation}
describing the uncorrelated part of the two-particle problem.
Thus we neglect in Eqs.(\ref{spectral})-(\ref{spectral-L}) the contribution of the transversal components (radiative corrections) and introduce in the longitudinal part the Yukawa potential (statically screened Debye potential) instead of the Coulomb potential to encompass the difficulties with the long-range interaction,
\begin{equation}
\label{eq:Yukawa}
    V_{\rm D}=\frac{1}{{\bf q}^2+m_{\rm D}^2}.
\end{equation}
Since our approach is focussed on the treatment of the interaction effects in the many-particle system, we consider the Yukawa interaction where the screening parameter $m_D$ is determined by the medium.

From a more sophisticated approach, we change from the $t$ channel to the $s$ channel for the interaction part.
The corresponding ladder diagram is also shown in the original paper of \cite{Baym:1962sx}.

The solution of the T-matrix equation \eqref{eq:T} corresponds to an in-medium wave equation, analogous to the Schr\"odinger equation, which contains corrections to the one-particle energies by self-energies, see Fig. \ref{fig:SE-T}. 
The Debye mass $m_{\rm D}$ is a medium effect which corresponds to the inverse screening length of the interaction, while phase space occupation effects (Pauli blocking) is contained in the intermediate one-frequency two-particle propagator $G_2^0(\Omega)$.
For details we refer to \cite{ZKKKR78} where excitons in the electron-hole semiconductor plasma are considered.  
In our case, the bound state is the positronium.

Coming back to the Beth-Uhlenbeck approach, we replace in Eq. \eqref{eq:OmegaPhi} the gauge boson propagator $D(\omega)$ by the two-particle propagator $G_2(\Omega)$, which has the same analytic properties as the two-particle T-matrix because of
\begin{equation}
    G_2(\Omega)\, V_{\rm D} = G_2^0(\Omega)\, T_2(\Omega),
\end{equation}
so that the derivations of Appendix B hold also for the replacement 
$D(\omega)\to G_2(\Omega) \to T_{2}(\Omega)$.
Analogous to the introduction of a phase in the polar representation of the complex gauge boson propagator in subsection 2.2, we can now introduce the scattering phase shifts corresponding to the two-particle T-matrix as
\begin{equation}
\label{eq:phase}
    \delta (\Omega) = \arctan\left[\frac{{\rm Im}\, T_2(\Omega)}{{\rm Re}\, T_2(\Omega)} \right].
\end{equation}
Following this analogy, we obtain that the contribution of the two-particle correlations (bound and scattering states) to the entropy assumes a form similar to Eq. \eqref{Sb-GBU}, 
\begin{equation}
    \label{S2-GBU}
	S_2 =
	2V\int\frac{{\rm d}^3P}{(2\pi)^3}~
%	\intbar_{0}^{\infty}
    \int_0^\infty \hspace{-5.5mm}-\hspace{3mm}
    \frac{{\rm d}\Omega}{\pi}~
	\sigma_n(\Omega)\, \sin^2\delta(\Omega,{\bf P})
	\frac{\partial \delta(\Omega,{\bf P})}{\partial \Omega} 
\end{equation}

We have shown that the $\sin^2\delta(\Omega)$ term is a necessary consequence and we conclude that also the scattering contribution gets modified.

The introduction of the generalized scattering phase shifts $\delta(\Omega,{\bf P})$ also takes into account the contribution of the bound states. 
This follows from the analytical form of $G_2(\Omega)$, which, in the low-density limit, takes the form
\begin{equation}
    G_2(\Omega)=\sum_{nP}|nP\rangle \frac{1}{E_{nP}-\mu_2-\Omega} \langle nP|,
\end{equation}
and $T_2(\Omega)$ also has the same analytical form. 
Here, $E_{nP}$ and $|nP\rangle$ are the eigenvalues and eigenstates of the in-medium  Schrödinger equation, where $P$ denotes the total momentum and angular momentum of the two-particle system (conserved quantities), while the intrinsic quantum number $n$ denotes both the bound states (discrete energy spectrum) and the scattering states (continuous energy spectrum), $\mu_2$ is the sum of the chemical potentials of the components.
The imaginary part of $G_2(\Omega)$ leads to $\delta$-like terms at the binding state energies.
Below the continuum threshold defined by the rest masses of the involved particles, a binding state causes a jump in the scattering phase shift by $\pi$.
According to the generalized Schrödinger equation discussed above, the binding energies are modified by the medium, and their disappearance defines the Mott effect. 

The modification of the Beth–Uhlenbeck formula -- specifically, the inclusion of the term $\sin^2 \delta (\Omega,{\bf P})$ and the modification of the few-particle states -- is often overlooked and ignored, which leads to incorrect results at high densities.  
}

%\begin{figure}[h]
%\centerline{a)\hspace{3mm}
%\includegraphics[width=0.1\textwidth]{Fig5A_cut.png}
%\hspace{2cm}
%b)\hspace{3mm}
%\includegraphics[width=0.1\textwidth]{5A_T-SE_bold.png}}
%\caption{\redflag{
%Feynman diagram for the self-energy.}
%\label{fig:T-SE}}
%\end{figure}

%%%%%%%%%%%%%%%%%%%%%%%%%%%%%%%%%%%%%%%%%%%%%%%%%%%%
\section{Discussion}
%%%%%%%%%%%%%%%%%%%%%%%%%%%%%%%%%%%%%%%%%%%%%%%%%%%%

The close relation between the conserving approximations in terms of $\Phi$,
and the consistent generalization of the Beth-Uhlenbeck formula is important because we need the introduction of  quasiparticles to cover a wide range of densities -- as known from mean-field approximations in Coulomb systems, relativistic mean-field approximations (RMF) in nuclear systems, and self-consistent (gap) equations in the quark-gluon plasma.
%\redflag
{In such approximations, correlations -- for example, in the Beth–Uhlenbeck formula -- must be handled carefully to avoid double counting.
This point has already been discussed in various articles such as \cite{ZKKKR78,Schmidt:1990oyr,Blaizot:2000fc,Blaschke:2016fdh}, but is often not taken into account in standard approaches to dense systems of interacting particles.
In particular, the Breit-Wigner formula is generally used to describe resonances in the continuum, which, as shown in Sect. \ref{sec:GBU}, is incorrect when a quasi-particle approach is used.}

The generalized Beth-Uhlenbeck  approach has the advantage over the standard one in that it is applicable also beyond the low-density limit, when due to phase space occupation, the spectral properties of correlations described by scattering phase shifts are strongly modified compared to the case of interactions in free space.
The quasiparticle picture is valid also in the high-density limit, where strongly correlated bound states are suppressed by Pauli blocking -- the Mott effect.
\iffalse
%\redflag
{We have presented here the simplest approximation for the self-energy contribution to the quasiparticle picture, the self-consistent Hartree-Fock approximation. 
If we consider higher approximations for the self-energy to determine the quasiparticle contribution in the generalized Beth-Uhlenbeck formula, the corresponding corrections have to be performed also for the two-particle contribution of this formula.}
\fi

In the strongly degenerate limit, the interacting fermion system is described as a Landau Fermi-liquid.
Importantly, the self-consistent mean-field (Hartree-Fock) approach discussed here
works not only at low densities but is also applicable as a self-consistently generalized Beth-Uhlenbeck formula at high densities.

%\redflag
{To give an example, in a recent application of the generalized Beth-Uhlenbeck approach to a (2+1)D Gross-Neveu model \cite{Mahato:2026kzr}, it has been shown that the $\sin^2\delta$ factor gives an important correction, in particular by suppressing the contribution of the low-energy Landau damping modes to the entropy.}
%\redflag{unklar:} For kinetic processes, it is relevant to determine how correlations are treated and how quasiparticles are formulated and included. The Boltzmann equation, in particular the Landau collision term, contributes to the entropy production.
%\\

%\redflag
{A related issue is the question of composition -- that is, what proportion of the particles are free or bound -- 
 such as the degree of ionization of a plasma.}
While the subdivision of the density into an 'free' quasiparticle part and the remaining correlated part is a formal issue, and in the final bookkeeping only the total density is relevant,  we have to avoid double counting, 
The discussion of free and bound density, ionization degree, etc. in plasma physics is an old problem, often discussed with ill-defined concepts. 
After separation of the single-quasiparticle term in the spectral function, the correlated contribution contains bound state contributions as well as contributions from the continuum. 

This concept was worked out for the proton-neutron system in Ref.~\cite{Schmidt:1990oyr}, where the self-consistent two-particle contribution contains a possible bound state (deuteron) and the correlated part from the continuum of scattering states.  Here empirical RMF expressions were used instead of the self-consistent mean-field shift of the quasiparticle energy; the Mott effect was obtained, and the contribution of scattering states was calculated using a separable potential approach.  One can introduce a generalized phase shift, as a function of temperature and density,
 to cover both the bound state and the scattering contribution.
 
An important issue in using the formalism described in our work is the question of  correlations higher than second order, 
for instance, if bound states are formed with more than two particles, such as molecules in a plasma.
We can formulate a cluster virial expansion \citep{Ropke:2012qv} in which atoms and ions are also treated as “elementary" particles.
The effective (Lennard-Jones) interaction leads to the formation of molecules. 
This chemical model should be embedded in a physical framework in which electrons and nuclei are regarded as fundamental elementary particles, which
can be achieved by introducing appropriate ladder diagrams. 
A systematic approach, valid also at high densities, has to be formulated consistently, including screening and Pauli blocking.

We mention as well the application of the Beth-Uhlenbeck approach to quark-hadron system, where three quarks can  bind to form a nucleon, and mesons are bound states of a quark and an antiquark, described as "plasmons" of the relativistic quark plasma within the Nambu--Jona-Lasinio model \cite{Hufner:1994ma,Zhuang:1994dw}. 
This model field theory is suitable for describing the dynamical chiral symmetry breaking transition which generates a quark mass gap in the vacuum and which leads to the pion being simultaneously the Goldstone boson of this broken symmetry and a bound state.  At finite temperatures and chemical potentials, restoration of chiral symmetry triggers the Mott dissociation of the pion which loses its character as a Goldstone boson and becomes a resonance in the quark plasma with a mass pole shifted to the complex energy plane, where the imaginary part of the 
self-energy describes the finite decay width (inverse lifetime) of this state. 
An achievement of the Beth-Uhlenbeck formulation with phase shifts has been a generalization to finite temperatures of the Levinson theorem for mesons \cite{Wergieluk:2012gd} and diquarks \cite{Blaschke:2013zaa,Blaschke:2014zsa}, 
as a prerequisite for describing baryons as quark-diquark bound states within a cluster virial expansion. The Beth-Uhlenbeck approach has recently been applied in Ref.  \cite{Mahato:2024fta} to the (2+1)D Gross-Neveu model as a model field theory to describe the Mott dissociation of excitons in graphene-type materials.
We note that phenomenological approaches that account for a finite lifetime of hadronic states in a resonance gas by simply introducing their spectral broadening violate the in-medium Levinson theorem! See also~\cite{zbp}.

In Ref. \cite{Blaschke:2016sqn} the Beth-Uhlenbeck approach to scalar-pseudoscalar mesons was extended to the three-flavor case with kaons and their chiral partner states. 
First extensions to the generalized Beth-Uhlenbeck scheme, the subject of the present work have been presented in %\cite{Blaschke:2015nma}
\cite{Blaschke:2016fdh}. 
This scheme has recently successfully described the transition from a full hadron resonance gas to the quark-gluon plasma, where Mott dissociation of hadrons and the simultaneous occurrence of quarks and gluons was driven by the chiral restoration transition \cite{Blaschke:2023pqd}. 
This work demonstrated that the description of the quark-hadron thermodynamics within the generalized Beth-Uhlenbeck-type model is in excellent agreement with data from {\em ab initio} lattice QCD thermodynamics and can be used to interpret them, e.g., by quantifying the composition of the system for given values of the  temperature and chemical potentials.

%\redflag
{
Our approach is related to the introduction of the quasiparticle concept. 
In standard approaches, the mean-field approximation assumes an uncorrelated medium, but correlations in the medium alter the properties of the quasiparticles. 
Of interest in this context is the Dyson–Bethe–Salpeter equation of motion for the two-particle Green’s function; see \cite{Schuck:2020kpj}.  
The kernel in Equation (4.3) in this report is divided into a static and a dynamic part, where
the static part, Equation (4.7) there, defines the mean-field contributions. 
This is also introduced by the cluster mean-field approach \citep{Ropke:1994,Jin:2016euh}. 
A strictly self-consistent treatment of correlations in dense systems, in which the correlations obtained from the solution of the in-medium Schr\"odinger equation also determine the self-energies of the quasiparticles, is a difficult problem for the future.}\\

 \section{Conclusions}\label{sec5}

We have shown here how a generalized Beth-Uhlenbeck equation of state, Eq. (\ref{Sb-GBU}), with the typical $\sin^2\delta(\omega)$ correction factor relative to the standard approach, can be obtained within the $\Phi-$derivable approach
to model quantum field theories,  assuming a two-loop $\Phi-$functional (yielding ``sunset" diagrams).
%\redflag
{This generalization is necessary, for example, to describe high-density systems in which mean-field effects must be taken into account. 
We also show that, instead of the commonly used Breit-Wigner shape often found to describe the contribution of resonances, a quadratic Lorentz profile must be used. 
This follows from the conservation laws underlying the $\Phi$-derivable approach.}

%\redflag
{The need to use the generalized Beth–Uhlenbeck equation of state also applies to the density $n$ as a function of $T$ and $\mu$; see Eq. (\ref{sn}). The common practice in the literature of neglecting the correction factor $\sin^2\delta(\omega)$ is not justified for dense systems, as soon as mean-field effects are taken into account in a quasiparticle approach.}
Such a choice is quite general and finds applications, e.g., in describing the Mott dissociation of ions in warm and dense matter, of nuclei in dense nuclear matter, or of hadrons in quark matter. 
%\redflag
{The generalization of the mean-field approach, which also accounts for correlations within the medium in a self-consistent manner, is the subject of future work.}

%\backmatter
\subsection*{Author contributions}

All authors have contributed to this text and agree with its publication.
%in CPP. 

\subsection*{Acknowledgments}
 With great pleasure we dedicate this paper to Michael Bonitz on his sixty-fifth birthday. He created the International Workshop series on ``Kadanoff-Baym Equations: Progress and Perspectives for Many-Body Physics"
%in Nonequilibrium Green's Functions" 
at the University of Rostock, where in the 1999 Edition the authors came together for the the first time in person and G.B. presented a fresh look on his $\Phi$- derivable approach from the perspective of the then just published Ref. 
\cite{Vanderheyden:1998ph} .

G.B. and D.B. are grateful to the WE-Heraeus Foundation for supporting their participation in the Seminar on "Optics and its Applications in Quantum Technologies" in Yerevan (Armenia), where the main idea for this contribution was conceived.   
The work of D.B. was part of the project No. 2021/43/P/ST2/03319 co-funded by the National Science Centre and the European Union Framework Programme for Research and Innovation Horizon 2020 under the Marie Skłodowska-Curie grant agreement No. 945339. 
For the purpose of Open Access, the author has applied a CC-BY
public copyright licence to any Author Accepted Manuscript (AAM) version arising from this submission.
G.R. acknowledges an honorary stipend from the Foundation for Polish Science within the Alexander von Humboldt program under grant No. DPN/JJL/402-4773/2022 which supports his visits at the University of Wroclaw where this work was finalized.

\subsection*{Financial disclosure}

None reported.

\subsection*{Conflict of interest}

The authors declare no potential conflict of interests.

%\bibliography{wileyNJD-AMA}
%\bibliographystyle{alpha}
%\bibliographystyle{plain}
%\bibliographystyle{unsrt}
%\bibliography{gbu.bib}

%\end{document}

%%%%%%%%%%%%%%%%%%%%%%%%%%%%%%%%%%%%%%%%%%%%%%%%%%%%
%\section*{References}
%%%%%%%%%%%%%%%%%%%%%%%%%%%%%%%%%%%%%%%%%%%%%%%%%%%%

%%%%%%%%%%%%%%%
% Appendices
%%%%%%%%%%%%%%%

\begin{appendix}

\section{Proving $S^\prime=0$ in scalar field theory with a $\phi^3$ self-interaction at two-loop order}
\label{app:warmup}
\begin{figure}[h]
\centerline{\includegraphics[width=0.1\textwidth]{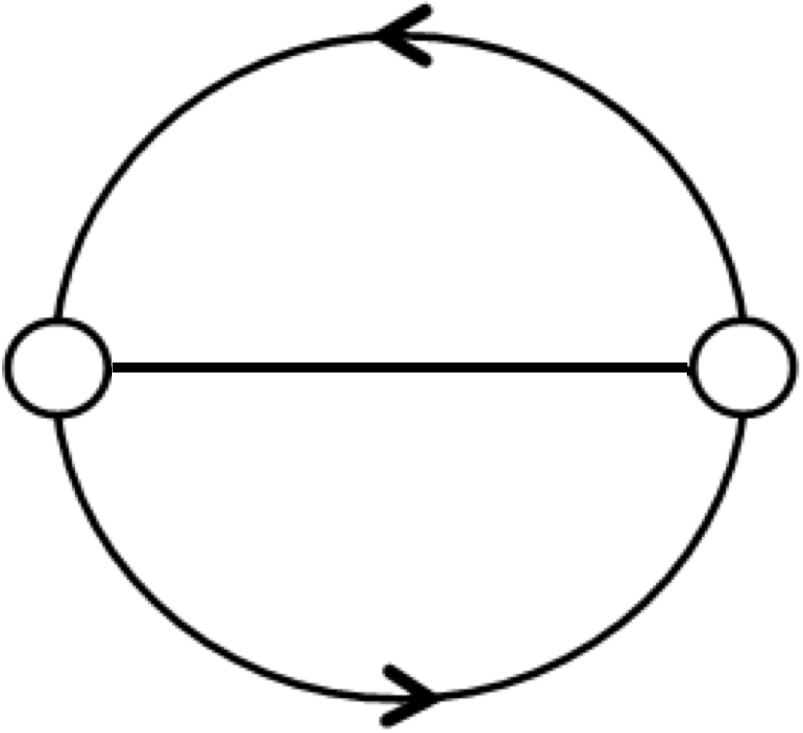}}
\caption{Feynman diagram for the $\Phi$-functional for scalar $\phi^3$ theory at two-loop order (sunset diagram).\label{fig:sunset}}
\end{figure}

\vspace*{12pt}
%\redflag{PLEASE REWRITE THIS APPENDIX INCLUDING WORDS, NOT SIMPLY FORMULAS!!!!!}
In this Appendix, we %\redflag
{provide a demonstration of} the cancellation expressed by the vanishing of $S^\prime$, Eq. \eqref{eq:Sprime}, for the simple case of a scalar field $\phi$ with a $\phi^3$ interaction vertex. 
%\redflag
{To this end we follow closely the calculation performed in \cite{Blaizot:2000fc}.}
The two-particle irreducible Feynman diagram for the $\Phi$-functional in two-loop approximation is shown in Fig. \ref{fig:sunset}.

 {
We write  $S^\prime$ for this case, 
\begin{equation}
    \mathcal{S}' \equiv - \frac{\partial (T \Phi/V)}{\partial T}\bigg|_{D} 
+ \int \frac{d^4k}{(2\pi)^4}\left\{\frac{\partial n(\omega)}{\partial T} {\rm Re} \Pi\, {\rm Im} D \right\},
\label{eq:A1}
\end{equation}
where the first term is
\begin{equation}
    - \frac{T}{V} \Phi = \frac{g^2}{12} T^2 \sum_{\omega_1,\omega_2} \int \frac{d^3k_1 d^3k_2}{(2\pi)^6}
D(\omega_1,|k_1|)D(\omega_2,|k_2|)D(-\omega_1-\omega_2,|-k_1-k_2|).
\end{equation}
Using the spectral representation
\begin{equation}
    D(\omega,k) = \int_{-\infty}^{\infty}\frac{dk_0}{2\pi} \frac{\rho (k_0,k)}{k_0-\omega}
\end{equation}
and evaluating Matsubara sums, we obtain
\begin{eqnarray}
    - \frac{T}{V} \Phi &=& \frac{g^2}{12} T^2 \sum_{\omega_1,\omega_2} \int \frac{d^4k d^4k' d^4k"}{(2\pi)^9}
\delta^{(3)}({\bf k+k'+k''})\rho(k)\rho(k')\rho(k'')
%\nonumber\\
\frac{-1}{\omega_1-k_0}\frac{-1}{\omega_2-k'_0}
\frac{1}{\omega_1+\omega_2+k''_0}~.
\end{eqnarray}
Partial fraction decomposition of the three energy denominators and Matsubara summation over $\omega_1$, $\omega_2$ yields
\begin{equation}
    \frac{1}{k_0+k'_0+k''_0}\left\{[n(k''_0)+1][n(k_0)+n(k''_0)+1] + n(k_0) n(k''_0) \right\}.
\end{equation}
Performing the temperature derivative and renaming variables under the integrals we obtain
\begin{eqnarray}
\partial_T\left[n(k_0+n(k'_0)+n(k''_0)+n(k'_0)n(k_0)+n(k'_0)n(k''_0)+n(k_0)n(k''_0) \right] 
%\nonumber\\
\to 3 \partial_T n(k_0)\left[1+n(k'_0)+n(k''_0) \right].
\end{eqnarray}
To evaluate the second term in Eq. \eqref{eq:A1}, we perform the Matsubara sums
\begin{eqnarray}
    {\rm Re} \Pi(\omega,q)&=&-\frac{g^2}{2}\int\frac{d^3k}{(2\pi)^3}\int\frac{dk_0}{2\pi}\int\frac{dk'_0}{2\pi}
\rho(k_0,|k|)\rho(k'_0,|{\bf k+q}|)
%\nonumber\\
\sum_{\omega_1}\frac{1}{\omega_1-k_0}~\frac{1}{\omega_1+\omega-k'_0}
\nonumber\\
&=&-\frac{g^2}{2}\int\frac{d^3k}{(2\pi)^3}\int\frac{dk_0}{2\pi}\int\frac{dk'_0}{2\pi}
\rho(k_0,|k|)\rho(k'_0,|{\bf k+q}|)
%\nonumber\\
\frac{1+n(k_0)+n(k'_0)}{\omega+k_0+k'_0}
\end{eqnarray}
and obtain 
}
\begin{eqnarray}
    \int\frac{d^4q}{(2\pi)^4} \frac{\partial n(k_0)}{\partial T} {\rm Re} \Pi(\omega,q) {\rm Im} D(\omega,q)
%    \nonumber\\
&=&-\frac{g^2}{2\cdot 2} \int\frac{d^4q}{(2\pi)^4}\int \frac{d^4k}{(2\pi)^4}\int \frac{d^4k'}{2\pi} 
\delta^{(3)}({\bf q+k+k'})\rho(q)\rho(k)\rho(k') 
\nonumber\\
&&\partial_Tn(q_0)\left[1+ n(k_0)+n(k'_0)\right] 
\frac{1}{q_0+k_0+k'_0}.
\end{eqnarray}
This proves the cancellation of $\mathcal{S}'$ for the scalar theory with cubic self-interaction in the 2-loop 
approximation (sunset diagram) for the $\Phi$-functional.
\\[5mm]
This cancellation holds as well for the density 
$n=N/V=-\partial \Omega/\partial \mu$, see \cite{Vanderheyden:1998ph}.

\section{Generalized optical theorems}
\label{app:GOT}
\vspace*{12pt}

In this Appendix, we prove the relation \eqref{eq:deropt} which is essential to derive the generalized Beth-Uhlenbeck formulas. 
The following steps are analogous to a similar derivation of generalized optical theorems for the thermodynamic T-matrix that was given in \cite{Zimmermann:1985ji}.
We denote the real and imaginary parts of complex functions 
$F(z,{\bf q})$ by subscripts $R$ and $I$, respectively, and suppress their arguments so that $F_I={\rm Im}\,F(\omega+i0^+,{\bf q})$ and $F_R={\rm Re}\,F(\omega,{\bf q})$.

We decompose the propagator and the photon self-energy (polarization function) into real and imaginary part, $D=D_R+i\, D_I$ and $\Pi=\Pi_R+i\,\Pi_I$, resp.
%where we use the notation $D_R={\rm Re} D(\omega,{\bf q})$, and $D_I={\rm Im} D(\omega+i\eta,{\bf q})$ for the propagator, and for the photon self-energy (polarization function).

Assuming that the inverse exists, we can write out the two identities 
$D=D^*{D^*}^{-1}D$ and $D^*=D^*D^{-1}D$ as,
\begin{eqnarray}
    	D_R+i\, D_I&=&D^*(D_R^{-1}-i\, D_I^{-1})D~,\\
	D_R-i\, D_I&=&D^*(D_R^{-1}+i\, D_I^{-1})D~.
\end{eqnarray}
Adding and subtracting these equations yields
\begin{eqnarray}
    	D_R&=&D^*D_R^{-1}D~,\\
	D_I&=&-D^*D_I^{-1}D~,
\label{SR}
\end{eqnarray}
which together with
$D^{-1}=D_0^{-1}-\Pi$, can be combined to
\begin{equation}
 %   	\label{OT}
	D_I=D^*\Pi_ID=D\Pi_ID^*~,
\label{SI}
\end{equation}
which is the off-shell optical theorem.

  Differentiating (\ref{SR}) both sides with respect to the energy $\omega$ 
(denoting this derivative with a prime) and using the fact, that in the Coulomb gauge the free photon propagator  
$D_0=1/{\bf q}^2$ is a real and independent of energy, so 
that $(D_R^{-1})^\prime = - \Pi_R^\prime$ and $D_I^{-1}=-\Pi_I$,  
we obtain
\begin{eqnarray}
    D_R^\prime &=& {D^*}^\prime D_R^{-1} D + D^* (D_R^{-1})^\prime  D + 
{D^*} D_R^{-1} D^\prime \nonumber \\
&=& {D^*}^\prime (\underbrace{D_R^{-1}+i\, D_I^{-1}}_{D^{-1}}- i\, D_I^{-1}) D 
    + D^* (D_R^{-1})^\prime  D + 
{D^*} (\underbrace{D_R^{-1} - i\, D_I^{-1}}_{{D^*}^{-1}} 
+ i\, D_I^{-1})D^\prime \nonumber \\
&=& \underbrace{{D^*}^\prime + D^\prime}_{2 D_R^\prime} 
    - i\,{D^*}^\prime D_I^{-1} D + i \,{D^*} D_I^{-1} D^\prime
    + D^* (D_R^{-1})^\prime  D \nonumber \\
&=& D^*\Pi_R^\prime D -i\, D^{*\prime}\Pi_ID +i\, D^*\Pi_ID^\prime
	~,
\label{dot}
\end{eqnarray}
which can be called a derivative optical theorem.

Multiplying  both sides of Eq.~(\ref{dot}) by $\Pi_I$ and using cyclic invariance of the product of matrices under the trace (integral over energy and momentum) and the optical theorem (\ref{SI}) one obtains the following.
\begin{equation}
    \label{derivativeOT}
D_R^\prime \,\Pi_I = \underbrace{D^*\,\Pi_R^\prime\, D\, \Pi_I}_{\Pi_R^\prime D_I}
	+ \underbrace{i\, D^*\,\Pi_I\,D^\prime\,\Pi_I-i\, D^{*\prime}\,\Pi_I\,D\,\Pi_I}_{
	2{\rm Im} \left[\Pi_ID\Pi_ID^{*\prime}\right]}
	~,
\end{equation}
which completes the proof of the relationship Eq. \eqref{eq:deropt} of the main text;
\begin{equation}
    \label{dot-rel}
D_R^\prime \,\Pi_I-\Pi_R^\prime\, D_I = 2{\rm Im} \left[\Pi_I\,D\,\Pi_I\,D^{*\prime}\right]~.
\end{equation}

\section{T-matrix and scattering phase shifts}%
\label{app:Tmatrix}
\vspace*{12pt}
In this Appendix, we present the derivation of the T-matrix expression for the generalized Beth-Uhlenbeck formula which parallels similar steps performed in the previous section for the full gauge boson propagator $D$.
While $D$ and the polarization function $\Pi$ fulfill the Dyson-Schwinger equation \eqref{eq:Dyson},
\begin{equation}
\label{eq:DSE}
    D^{-1}=D_0^{-1}-\Pi,
\end{equation}
the T-matrix obeys the Bethe-Salpeter equation \eqref{eq:T} in the ladder approximation, which can be rewritten in the form
\begin{equation}
\label{eq:BSE}
    T_2^{-1}=V_D^{-1}-G_2^0.
\end{equation}
Using the formal equivalence of  \eqref{eq:BSE} with \eqref{eq:DSE} with the dictionary:
$D\to T_2$, $D_0\to V_D$ and $\Pi \to G_2^0$, we can simply rewrite Eq. \eqref{sin2} now for the T-matrix,
\begin{equation}
    \label{eq:DMB}
    \left[{\rm Re}\,T_2\, {\rm Im}\,G_2^0 + {\rm Im}\,\ln(-T_2^{-1})\right]^\prime
= - 2{\rm Im}\left({\rm Im}\,G_2^0\, T\, {\rm Im}\,G_2^0\,  {T^*}^\prime\right) ,
%\nonumber\\ &=& 2 \delta^\prime \sin^2\delta.
\end{equation}
where the prime denotes the derivative with respect to the energy:
$F^\prime=dF/d\omega$.
Here we recognize in \eqref{eq:DMB} the generalization of the corresponding expression in 
\cite{Dashen:1969ep}, now for the high-density case where in-medium effects are accounted for by the phase space occupation factor (Pauli blocking) and the one-particle self-energy contained in ${\rm Im}G_2^0$.
Introducing the scattering phase shift \eqref{eq:phase} of the two-particle T-matrix by its polar representation $T_2=|T_2|\exp{(i\delta)}$,
so that 
${\rm Im}G_2^0={\rm Im}(V_D^{-1}-T_2^{-1})=|T_2|^{-1}\,\sin\delta $,
because the Yukawa potential $V_D$ is a real quantity, we can rewrite the expression \eqref{eq:DMB} as 
\begin{equation}
    \label{eq:sin2}
%    \left[{\rm Re}T_2\, {\rm Im}G_2^0 + {\rm Im}\ln(-T_2^{-1})\right]^\prime &=& 
- 2{\rm Im}\left({\rm Im}G_2^0\, T\, {\rm Im}G_2^0\, {T^*}^\prime\right) 
%\nonumber\\ 
= 2 \delta^\prime \sin^2\delta.
\end{equation}
With this expression we recover the generalized Beth-Uhlenbeck formula \eqref{Sb-GBU} for the entropy contribution from two-particle correlations, where $\delta$ is now the scattering phase shift \eqref{eq:phase} of the two-particle T-matrix that fulfills the Bethe-Salpeter equation in the ladder approximation with a two-particle potential as, e.g., the Yukawa potential \eqref{eq:Yukawa} that accounts for screening effects by the medium dependent Debye mass $m_D$.

\end{appendix}

\end{document}